\documentclass[11pt,a4paper]{article}
\usepackage{jinstpub}
\usepackage{textgreek}
\usepackage{siunitx}
\usepackage{isotope}
\usepackage{csquotes}

\title{A thermionic electron gun to characterize silicon drift detectors with electrons}

\author[a,b,1]{K.~Urban\note{Corresponding author.},}
\author[c]{M. Biassoni,}
\author[d,e]{M. Carminati,}
\author[a,b]{F. Edzards,}
\author[d,e]{C. Fiorini,}
\author[a,b]{C. Forstner,}
\author[f]{P. Lechner,}
\author[c,g]{A. Nava,}
\author[a,b]{D. Siegmann,}
\author[a,b]{D. Spreng}
\author[a,b]{and S. Mertens}

\affiliation[a]{Technical University of Munich, TUM School of Natural Sciences, Physics Department, James-Franck-Str. 1, 85748 Garching, Germany}
\affiliation[b]{Max Planck Institute for Physics, Boltzmannstr. 8, 85748 Garching, Germany}
\affiliation[c]{INFN, Sezione di Milano-Bicocca, Piazza della Scienza 3, 20126 Milano, Italy}
\affiliation[d]{INFN, Sezione di Milano, Via Celoria 16, 20133 Milano, Italy}
\affiliation[e]{DEIB, Politecnico di Milano, Via Golgi 40, 20133 Milano, Italy}
\affiliation[f]{MPG Semiconductor Laboratory, Isarauenweg 1, 85748 Garching, Germany}
\affiliation[g]{University of Milano-Bicocca, Piazza della Scienza 3, 20126 Milano, Italy}

\emailAdd{korbinian.urban@tum.de}

\abstract{
The TRISTAN detector is a new detector for electron spectroscopy at the Karlsruhe Tritium Neutrino~(KATRIN) experiment.
The semiconductor detector utilizes the silicon drift detector technology and will enable the precise measurement of the entire tritium \textbeta-decay electron spectrum.
Thus, a significant fraction of the parameter space of potential neutrino mass eigenstates in the \si{keV}-mass regime can be probed.
We developed a custom electron gun based on the effect of thermionic emission to characterize the TRISTAN detector modules with mono-energetic electrons before installation into the KATRIN beamline.
The electron gun provides an electron beam with up to \SI{25}{keV} kinetic energy and an electron rate in the order of \num{E5} electrons per second.
This manuscript gives an overview of the design and commissioning of the electron gun.
In addition, we will shortly discuss a first measurement with the electron gun to characterize the electron response of the TRISTAN detector.}

\begin{document}
\maketitle
\section{Introduction}
The sterile neutrino program of the Karlsruhe Tritium Neutrino~(KATRIN) experiment searches for a kink-like distortion in the tritium \textbeta-decay spectrum due to a heavy neutrino mass eigenstate~\cite{aker_22}.
Two unknown parameters characterize the parameter space of a potential heavy mass eigenstate:
its mass $m_4$ and its mixing amplitude $\sin^2\theta$ to the electron neutrino flavor eigenstate.
Currently, the KATRIN experiment is optimized to measure the tritium \textbeta-decay spectrum in the last \SI{40}{eV} below the spectral endpoint $E_0=\SI{18.6}{keV}$~\cite{aker_22b}.
As the kink signature occurs at $E = E_0 - m_4c^2$, this energy range limits the search to sterile neutrinos to masses $m_4<\SI{40}{eV/c^2}$.
To probe heavier masses, the energy range has to be extended.
During the commissioning phase of the KATRIN experiment, a test measurement with reduced source strength and extended measurement interval of \SI{1.6}{keV} was performed~\cite{aker_23}.
This measurement probed mass eigenstates up to $m_4\approx\SI{1.6}{keV/c^2}$ and mixing amplitude down to $\sin^2\theta\approx \num{5E-4}$.
To expand the search to the entire phase space of tritium \textbeta-decay and to mixing angles down to $\sin^2\theta\approx\num{E-6}$, the energy interval and the number of collected electrons must be increased.
To this end, a new detector system, the TRISTAN detector, is currently being developed~\cite{mertens_19}.
The new detector system will allow for measuring the electron energy spectrum at high total count rates of about \num{E8} counts per second~(cps).

The TRISTAN detector is implemented as a silicon drift detector~(SDD) array.
The low capacitance of the anode at the center of the pixel reduces the noise contribution from the amplification circuit~\cite{lechner_01}.
This enables an energy resolution close to the fundamental Fano limit (\SI{216}{eV} FWHM at \SI{20}{keV}) of silicon detectors.
The TRISTAN detector is a multi-pixel focal plane array of at least nine detector modules of size ${38\times 40~\si{mm^2}}$, each consisting of 166 hexagonal pixels of \SI{7}{\mm\squared} each.
The analog low-noise readout of the detector uses a dedicated application-specific integrated circuit~(ASIC) called ETTORE~\cite{trigilio_18,gugiatti_22, carminati_23}.

One of the main challenges of the TRISTAN detector is the application of the SDD technology to high-precision electron spectroscopy.
The \textbeta-electrons will hit the detector with a maximal kinetic energy of \SI{38.6}{keV}, which is the sum of the spectral endpoint ${E_0=\SI{18.6}{keV}}$ and an optional accelerating potential of up to \SI{20}{kV}.
The electrons deposit their kinetic energy in the detector by scattering along a path of several \textmu m in the detector volume, starting at the entrance window surface.
The properties of charge collection and scattering at the entrance window substantially impact the detector's response to electrons.

In the past, the electron-related effects for SDDs were measured and compared to electron tracking simulations~\cite{gugiatti_20,biassoni_21,nava_21,mertens_21}.
To measure the detector response to mono-energetic electrons in the laboratory, two kinds of sources were used: 1) the beam of a scanning electron microscope, and 2) radioactive sources such as \isotope[83]{Rb} or \isotope[109]{Cd} providing mono-energetic conversion electrons from nuclear transitions.
However, a scanning electron microscope cannot be used to characterize the 166-pixel TRISTAN detector module because the detector and readout electronics do not fit into the vacuum chamber of the microscope. Previous measurements were possible only for prototype detectors with up to 12 pixels.
Moreover, using electrons from radioactive sources has several drawbacks: The emitted line spectra are distorted due to the non-avoidable scattering of the electrons in the source material. Furthermore, the fixed rate and fixed line position limit the full characterization of the detector.
This work presents a dedicated electron source (called electron gun in the following) that overcomes these restrictions.
The key requirements of the electron gun are to
\begin{enumerate}
    \item provide mono-energetic electrons with a kinetic energy up to \SI{25}{keV},
    \item allow electron rates up to \SI{100}{kcps}, and
    \item illuminate the area of one TRISTAN detector module ($38\times\SI{40}{mm^2}$).
\end{enumerate}

The electron gun presented in this work differs from the photo-electron source, which sits at the rear section of the KATRIN beamline.
The characteristic of the KATRIN rear section electron source is its exact energy and low angular spread, which is necessary to calibrate the transmission properties of the KATRIN spectrometer~\cite{behrens_17}.
Compared to that, the electron gun presented in this work has less stringent requirements for the energy spread of the electrons.
Furthermore, it operates outside the KATRIN beamline in a flexible and accessible setup without superconducting magnets.  
The new electron gun is intended to be used as a flexible tool for many applications. These applications include detailed characterization measurements with a representative TRISTAN detector module and the verification of the performance of each single detector module after assembly.

\section{Electron gun design}\label{sec:electron_gun_design}

\begin{figure}
    \centering
    \includegraphics{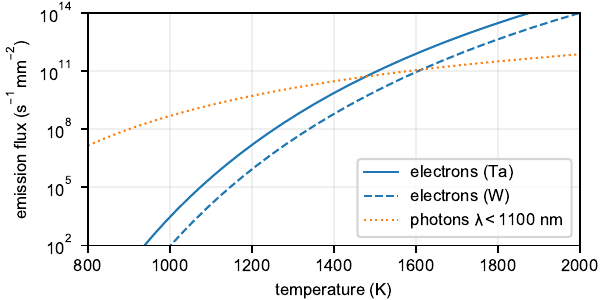}
    \caption{Calculated emission flux of electrons as function of surface temperature according to Richardson's law (equation~\ref{eq:richardson}). Work functions values of $E_w=\SI{4.25}{eV}$ and $E_w=\SI{4.55}{eV}$ are used for tantalum (Ta) and tungsten (W), respectively~\cite{haynes_14}. The unfavorable emission flux of light (photons) is  also shown. The photon flux is calculated by integrating Planck's law of the blackbody spectrum for wavelengths $\lambda<\SI{1100}{nm}$. In this wavelength region, a silicon detector is sensitive to photons. The absolute scale of the curves is not essential. The photon line is multiplied by $10^{-9}$ to allow easier comparison.}
    \label{fig:emission}
\end{figure}

The electron gun designed within the scope of this work is based on thermionic electron emission~\cite{iqbal_05}.
Compared to other mechanisms used to generate free electrons, such as photo emission or high field emission, this approach has the advantage of a simple setup and the capability of providing almost unlimited electron rates.
The current density $J$ of electrons from the thermionic emission of a hot surface is estimated with Richardson's law
\begin{equation}\label{eq:richardson}
    J = A T^2\exp{\left(-\frac{E_w}{k_BT}\right)}\,.
\end{equation}
Here, $A\approx\SI{1.2E6}{Am^{-2}K^{-2}}$ denotes a constant, $T$ the temperature, $E_w$ the work function, and $k_B$ the Boltzmann constant~\cite{fowler_28}.
Given the work function of tantalum with a value of $E_w=\SI{4.25}{eV}$~\cite{haynes_14}, the electron flux changes by almost eight orders of magnitude between $T=\SI{1000}{K}$ and $T=\SI{1500}{K}$, see figure~\ref{fig:emission}.
Compared to the usual ambit of thermionic electron guns (for example, electron microscopy, cathode ray tubes, etc.), the electron rate required to characterize the TRISTAN detector is very small.
The targeted rate of \SI{100}{kcps} corresponds to a current of only \SI{0.016}{pA}.

\begin{figure}
    \centering
    \includegraphics{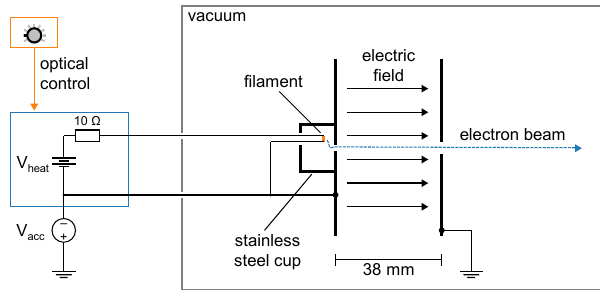}
    \caption{Schematic of the electron gun. The filament is inside a stainless steel cup to shield light from glowing. The position of the filament is around \SI{3}{mm} off-axis to have no direct line-of-sight to the detector. The electric field between two plates on $V_\text{acc}$ and ground accelerates the electrons from the filament to form an electron beam. Outside of the vacuum chamber is a high voltage power supply $V_\text{acc}$, a battery box (blue box) and a galvanically decoupled controller (orange box) to adjust the voltage $V_\text{heat}$.}
    \label{fig:electronic_sketch}
\end{figure}

\begin{figure}
    \centering
    \includegraphics{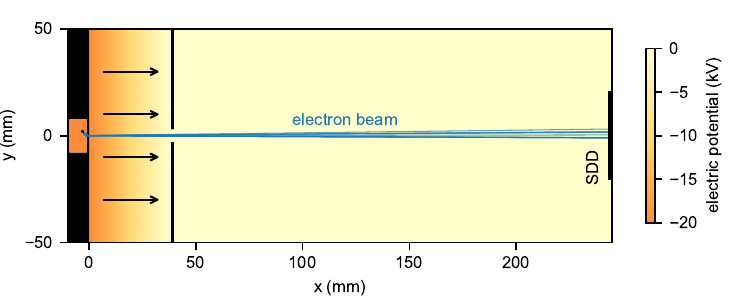}
    \caption{Simulation of the electric potential and ten electron tracks (blue) originating at the hot filament (small dot). The acceleration voltage is $V_\text{acc}=\SI{20}{kV}$. The figure shows only those tracks that reach the acceleration region. The spot size at the detector has a diameter of around \SI{4}{mm}.}
    \label{fig:efield_sim}
\end{figure}

A schematic of the electron gun is shown in figure~\ref{fig:electronic_sketch}.
A thin tantalum wire (in the following called filament) with a length of \SI{5}{mm} and a diameter of \SI{25}{\micro\meter} is heated directly by a current of around \SI{150}{mA} to reach temperatures above \SI{1000}{K}.
The filament is assembled in a stainless steel cup on negative high voltage.
The stainless steel cup shields the light from the glowing filament.
The free electrons can leave the cup and reach the electric field region through a hole of \SI{0.5}{mm} diameter.
A plate capacitor with a plate distance of \SI{38}{mm} forms the acceleration field.
The electric field and several electron trajectories have been simulated to verify this design.
The field simulation uses the \textit{sfepy} software, a finite element calculation package for python~\cite{cimrman_19}.
The electric potential and electron tracks in the electron gun are shown in figure~\ref{fig:efield_sim}.
The resulting beam spot is smaller than the sensitive detector area ($38\times\SI{40}{mm^2}$).
The exact size of the beam spot depends strongly on the geometry in the filament region and the acceleration voltage.
It will, therefore, be determined experimentally.

\subsection{Light emission}
When operating a glowing cathode in the vicinity of a silicon detector, the light emission disturbs the detector.
The constant illumination of the detector diode with light leads to a photocurrent, which adds to the diode leakage current and worsens the energy resolution.
To reduce the number of detected photons, three measures have been applied in the design of the electron gun:
\begin{itemize}
    \item The material of the filament is tantalum~(Ta) with a work function of $E_w=\SI{4.25}{eV}$~\cite{haynes_14}. Compared to tungsten~(W) with a work function of $E_w=\SI{4.55}{eV}$, this lowers the required temperature by around \SI{100}{K}, see figure~\ref{fig:emission}. Other materials with even lower work functions, like oxid-coated cathodes, are usually unstable in air, making the handling impractical. 
    \item Figure~\ref{fig:emission} shows that the ratio of electron-to-photon emission is more favorable for higher temperatures.
    This motivates a small emission area at high temperatures.
    Therefore, a thin filament with a diameter of \SI{25}{\micro\meter} was chosen.
    \item A considerable reduction of the number of photons that arrive at the detector is achieved by placing the filament around \SI{3}{mm} off-axis to have no direct line-of-sight to the detector.
\end{itemize}

\subsection{Steering coils system} 
The tracking simulation in figure~\ref{fig:efield_sim} shows that the beam spot of the electron gun is smaller than the sensitive detector area ($40\times\SI{38}{mm^2}$).
For many characterization measurements, however, it is essential to illuminate the sensitive detector area homogeneously with electrons.
To this end, an electron lens could increase the beam divergence further, but this option has yet to be tested.
Another possibility to illuminate the entire detector area is to use a steering coil system presented in this work.
The steering coil system is placed near the electron source and allows for a magnetic deflection of the electron beam up to several degrees.
The coil system consists of two coils with 80 turns each.
Using a coil current of \SI{1}{A}, a magnetic field perpendicular to the beam of $B_{\perp}=\SI{2.8}{mT}$ and a bending power of $\int{B_{\perp}\rm{d}s}=\SI{6.7E-5}{Tm}$ is achieved.
The deflection angle $\theta$ can be estimated via
\begin{equation}
    \sin(\theta)=\frac{v_{\perp}}{v}\approx\frac{e\int{B_{\perp}\rm{d}s}}{\sqrt{2mE}}.
\end{equation}
Here, $v_{\perp}$ denotes the electron velocity component perpendicular to the beam and $v$ the total velocity.
Moreover, $E$ is the kinetic energy, and $m$ is the mass of the electron.
For $E=\SI{20}{keV}$, a deflection angle of $\theta=\SI{8.1}{\degree}$ is obtained, which is sufficient to reach the edges of the SDD chip at a distance of \SI{15}{cm}. 

\begin{figure}
    \centering
    \includegraphics{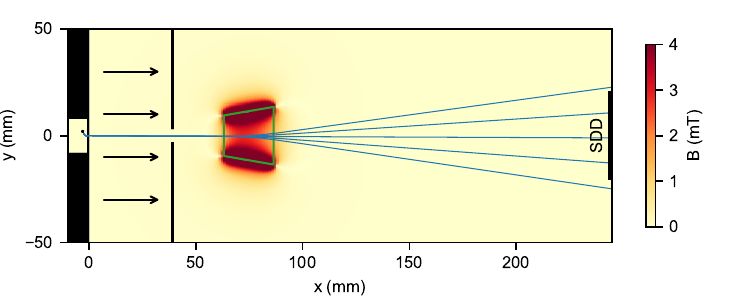}
    \caption{Simulation of the magnetic field of the steering coil for vertical deflection.
    The top view of the coil path is drawn in green. The five electron tracks correspond to a coil current of $I=\SI{-1}{A}$, $\SI{-0.5}{A}$, $\SI{0}{A}$, $\SI{0.5}{A}$, and $\SI{1.0}{A}$. The acceleration voltage is $V_\text{acc}=\SI{20}{kV}$. With this range of the coil current, the electrons can reach the entire detector area of the SDD.}
    \label{fig:steering_sim}
\end{figure}

Figure~\ref{fig:steering_sim} shows a simulation of the magnetic field and an electron trajectory at five different coil currents.
The entire detector area is reached by adjusting the currents of the two coils for horizontal and vertical deflection.
Applying triangular signals with a high frequency to the steering coils imitates a homogeneous illumination of the entrance window.
One should mention, however, that in this case, the scanning frequency (typically \SI{2}{kHz}) modulates the arrival times of the electrons on a single pixel.
This modulation must be considered when studying timing-related effects like event pileup.

\subsection{Experimental setup}
\begin{figure}
    \centering
    \includegraphics{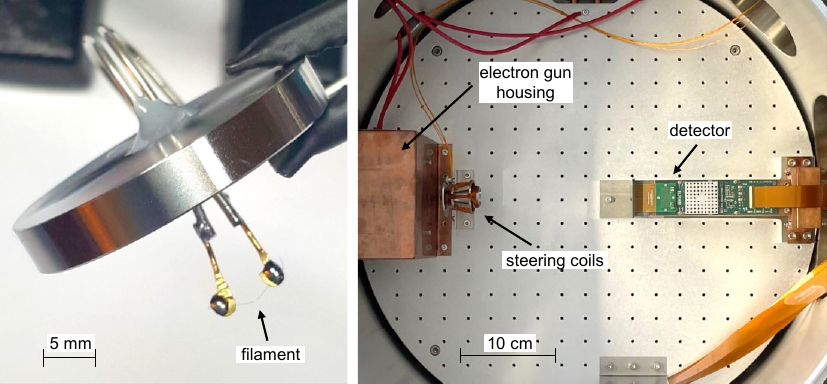}
    \caption{Photographs of the filament~(left) and of the measurement setup~(right).}
    \label{fig:filament_setup}
\end{figure}

The measurement setup to characterize the TRISTAN detector module with the electron gun consists of a cylindrical vacuum chamber with a liquid-cooled baseplate, see figure~\ref{fig:filament_setup}.
The liquid-cooled baseplate allows for the cooling of the detector chip to around \SI{-30}{\celsius}.
A turbomolecular pump evacuates the chamber.
Pressure levels on the order of ${\SI{E-6}{mbar}}$ can be reached after \SI{4}{h} of pumping.
The filament's electric supply is a battery biased to negative high voltage.
The heating voltage $V_{\rm{heat}}$ is applied to the filament via a series resistor (\SI{10}{\ohm}) and two high voltage vacuum feedthroughs as illustrated in figure~\ref{fig:electronic_sketch}.
A high-voltage optocoupler circuit is used so that the user can set the value of $V_{\rm{heat}}$ from ground potential.
The acceleration voltage $V_{\rm{acc}}$ is set by an adjustable precision high voltage supply with negative polarity.
The steering coils are driven by a custom-designed dual-channel power amplifier, similar to an audio amplifier.
Two signal generator channels are used to set the currents for the horizontal and vertical deflection.
The TRISTAN detector chip (166 pixels) is biased and read out by two dedicated front-end electronics boards~\cite{gugiatti_22} together with a bias and control board.
Each front-end board hosts seven 12-channel ETTORE ASICs.
The data acquisition system consists of three synchronized CAEN VX2740 digitizer cards\footnote{\href{https://www.caen.it/products/vx2740/}{www.caen.it/products/vx2740/}}. The cards feature full waveform digitization and online pulse-height analysis.
For the interconnection between inside-vacuum electronics and outside-vacuum electronics, four 100-pin Micro-D connectors are used.

\section{Commissioning}
Several commissioning measurements were performed to characterize the electron gun.
First, we present the measurement of the filament's voltage-current curve and its light emission.
Subsequently, measurements with electrons at acceleration voltages up to $V_\text{acc}=\SI{25}{kV}$ are presented.
These comprise tests of the steering coil system and rate stability.

\subsection{Voltage-current curve of the filament}
\begin{figure}
    \centering
    \includegraphics{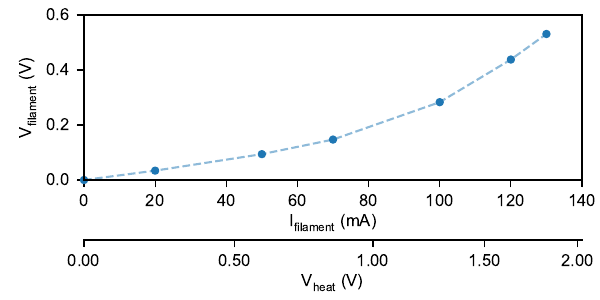}
    \caption{Measured voltage-current curve of the tantalum filament.
    Without heating, the resistance is \SI{1.7}{\ohm}.
    When heating the filament with a current of \SI{130}{mA}, the temperature increases, and the resistance changes to \SI{4.1}{\ohm}. On the second x-axis, the corresponding value of $V_{\rm{heat}}$ is shown. $V_{\rm{heat}}$ is the voltage including a \SI{10}{\ohm} series resistor, see figure~\ref{fig:electronic_sketch}.}
    \label{fig:ui_curve}
\end{figure}

In the first step, a measurement of the filament voltage versus current
was performed.
For this measurement, the acceleration voltage and the detector were switched off.
The result is shown in figure~\ref{fig:ui_curve}.
Due to the positive temperature coefficient of the electric resistivity of tantalum, a non-linear behavior is observed.
To adjust the electron rate of the electron source, the heating voltage $V_\text{heat}$ is regulated.
This changes the filament current and thus its dissipated power and filament temperature.

\subsection{Impact of light emission}
\begin{figure}
    \centering
    \includegraphics{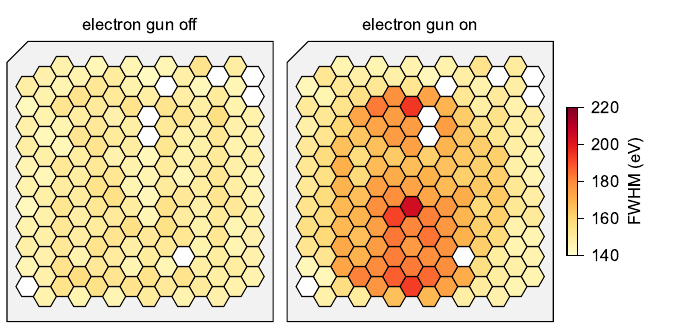}
    \caption{Full width at half maximum (FWHM) energy resolution of the $\text{Mn K}_\alpha$ X-ray peak of an \isotope[55]{Fe} calibration source, shown for all pixels of a 166-pixel SDD focal plane array. For the \textquote{electron gun on} measurement, the filament was heated with a heating voltage of $V_{\rm{heat}}=\SI{1.85}{V}$. The acceleration voltage $V_\text{acc}$ was zero to be sensitive only to the light emission effect. Eight pixels of the detector (shown in white color) had to be disabled due to connection issues. The risetime of the trapezoidal energy filter was set to \SI{2}{\micro s}.}
    \label{fig:pm_fwhm_calib}
\end{figure}

In the next step, the impact of the light emission of the electron gun on the detector performance was investigated.
To this end, the energy resolution of the TRISTAN detector was measured using an \isotope[55]{Fe} X-ray calibration source.
For this measurement, the acceleration voltage $V_\text{acc}$ was set to zero so that no electrons from the electron gun are expected to arrive at the detector.
One reference measurement of the \isotope[55]{Fe} X-ray energy spectrum was conducted, and another measurement was performed with additional heating of the filament.
The energy resolution of the two measurements is illustrated in the pixel maps in figure~\ref{fig:pm_fwhm_calib}.
An excellent mean energy resolution of \SI{150}{eV} full width at half maximum (FWHM) for the $\text{Mn-K}_\alpha$ peak at $E=\SI{5.9}{keV}$ is observed very homogeneously over all pixels in the reference measurement.
When a heating voltage is applied to the filament, the energy resolution worsens to values up to \SI{207}{eV} FWHM in some pixels of the detector. The observed pattern results from the light reflection through the hole in the electron gun plate and the steering coil system.

The electronics noise of a measurement system is typically quantified in terms of the equivalent noise charge~(ENC), which is expressed in units electrons root mean square (\si{el_{rms}}).
The energy resolution expressed in terms of FWHM and the ENC are related via the following equation~\cite{kolanoski_20}:
\begin{equation}\label{eq:fwhm}
    \text{FWHM}=\sqrt{2\log{2}}\cdot\sqrt{\sigma_\text{Fano}^2 + \text{ENC}^2\epsilon^2}.
\end{equation}
In this equation, $\epsilon=\SI{3.62}{eV}$ describes the average pair creation energy of silicon and $\sigma_\text{Fano}$ is the Fano energy resolution according to 
\begin{equation}\label{eq:fano}
    \sigma_\text{Fano}^2=0.115\epsilon E.
\end{equation}
The resulting electronics noise is $\text{ENC}\approx \SI{11}{el_{rms}}$ for the measurement with the electron gun switched off, and $\text{ENC}\approx \SI{20}{el_{rms}}$ for the most-affected pixel with the electron gun switched on.
At  $E=\SI{20}{keV}$, the energy resolution according to equation \ref{eq:fwhm} is  dominated by the Fano energy resolution.
The worsening due to the increased ENC, caused by residual light of the electron gun, is \SI{17}{\percent}.
This value is acceptable for TRISTAN detector characterization measurements.
Nevertheless, the effect of light must be considered when interpreting the energy resolution measured with the electron gun.

\subsection{Test of steering coil system} 
\begin{figure}
    \centering
    \includegraphics{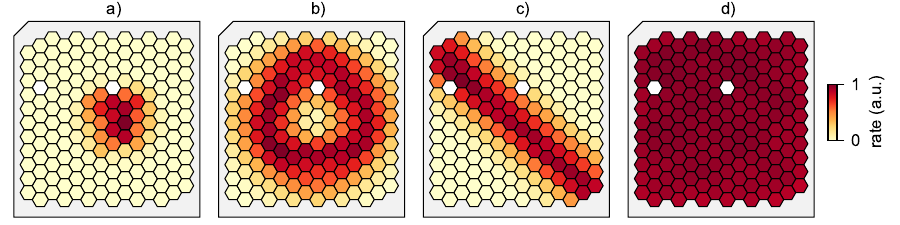}
    \caption{Illumination of the entrance window of a 166-pixel SDD focal-plane array with electrons using four different patterns. Pixel map a) shows the count rate of electrons if the steering coils are switched off. Pixel map b) shows the rate distribution for an acquisition where two sinusoidal signals with \SI{90}{\degree} phase shift were applied to the steering coils to draw a circle. Acquisition c) had two synchronous triangular signals applied to draw a diagonal line. Acquisition d) had two non-synchronous triangular signals with different frequencies applied, leading to a homogeneous illumination of the detector chip. Two pixels of the detector (shown in white color) had to be disabled due to connection issues.}
    \label{fig:pm_rates}
\end{figure}

The steering coil system guides the electron beam to any position on the detector surface.
Figure~\ref{fig:pm_rates} compares the rate distribution, measured by the 166-pixel detector, for four different signals applied to the steering coils.
If the steering coils are switched off, an electron beam spot with a size of around \SI{10}{mm} FWHM is observed at the center of the chip.
Patterns such as a circle and a line can be drawn on the detector surface by applying synchronous signals to both coils.
Finally, if two triangular signals with different frequencies are applied, the entire surface of the detector is illuminated.
A rate variation of about \SI{12}{\percent} was observed between the pixels for the acquisition with a homogeneous illumination.

\subsection{Electron rate}\label{sec:electron_rate}
\begin{figure}
    \centering
    \includegraphics{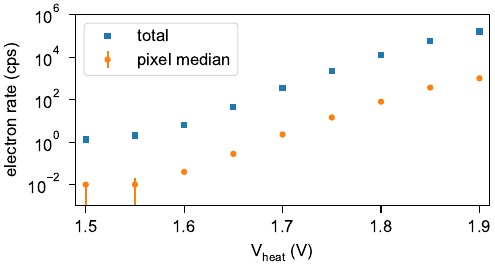}
    \caption{Measurement of the electron rate as a function of the heating voltage $V_\text{heat}$.
    The steering coil system was used to illuminate the entire detector array. The acceleration voltage was $V_\text{acc}=\SI{15}{kV}$.}
    \label{fig:el_rate}
\end{figure}

The rate of the electron gun is controlled by setting the heating voltage~$V_\text{heat}$.
Figure~\ref{fig:el_rate} shows how the detected rate depends on $V_\text{heat}$.
The rate was calculated by summing all events above \SI{8}{keV} to exclude events from the \isotope[55]{Fe} source ($E<\SI{6.5}{keV}$), which was installed in the setup for calibration purposes.
At $V_\text{heat}=\SI{1.5}{V}$, no electrons from the electron source are observed. A background rate of \SI{1.5}{cps} is present. The background originates from the pileup of two consecutive \isotope[55]{Fe} events.
At $V_\text{heat}\approx \SI{1.55}{V}$, the rate starts to increase due to electrons from the electron gun.
The electron source reaches a total rate of about \SI{160}{kcps} at $V_\text{heat}=\SI{1.9}{V}$.

The total rate of \SI{160}{kcps} is sufficient for almost all characterization purposes.
However, the rate could be increased further by raising $V_\text{heat}$.
To mitigate the excessive light on the detector at high rates, geometric modifications, such as a magnetic cascade, could be considered.
The optimal geometry for high-rate operation well above \SI{160}{kcps} remains to be determined experimentally.

\subsection{Rate stability} 
\begin{figure}
    \centering
    \includegraphics{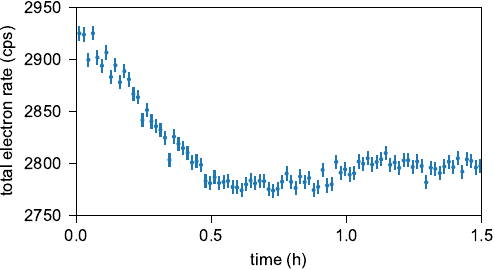}
    \caption{Measurement of the temporal evolution of the electron rate. The steering coil system was used to illuminate the entire detector area. The acceleration voltage was $V_\text{acc}=\SI{20}{kV}$. The electron gun was switched on \SI{1}{h} before the measurement so that the temperature of the filament could stabilize.}
    \label{fig:stability}
\end{figure}

Several effects can cause instabilities of the detected electron rate from the electron gun over time.
A change of the filament temperature by \SI{1}{K} at \SI{1300}{K} corresponds to a change in the rate of emitted electrons of approximately \SI{3}{\percent}, see figure~\ref{fig:emission}.
Therefore, a change in rate of several percent on the timescale of hours is expected.
Furthermore, any electric field in the vacuum chamber influences the trajectories of the electrons from the filament to the detector.
Although most parts of the setup were fabricated of metal and are electrically grounded, some isolated surfaces and cables could charge up and change the electric field over time.
A measurement of the electron rate over a period of \SI{1.5}{h} after a  is shown in figure~\ref{fig:stability}.
In the first \SI{0.5}{h} the rate dropped by \SI{5}{\percent}.
In the subsequent \SI{1}{h} period, a stability of \SI{1.3}{\percent} was observed.
This hints at a more stable rate for a longer measurement duration, since the electron gun has more time to stabilize.
A rate stability of \SI{5}{\percent} of the electron gun is sufficient for the characterization measurements of the TRISTAN detector system, as the focus of the detector characterization is the spectral detector response to electrons.

\section{Detector characterization}
\begin{figure}
    \centering
    \includegraphics{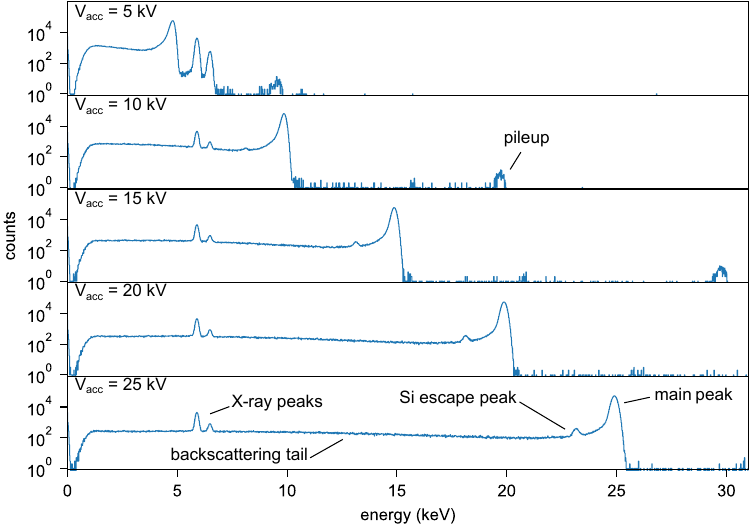}
    \caption{Measurement of the electron energy spectrum at different acceleration voltages $V_\text{acc}$.
    The spectra are recorded with a single pixel. The electron rate is \SI{1}{kcps} and the measurement duration is 20~minutes.
    For a precise energy calibration, an \isotope[55]{Fe} X-ray source is used in parallel to the electron gun, providing X-rays at \SI{5.9}{keV} and \SI{6.5}{keV}. The risetime of the trapezoidal energy filter was set to \SI{2}{\micro s}.}
    \label{fig:el_spectra}
\end{figure}

The detector response to mono-energetic electrons was measured at five different acceleration voltages: ${V_\text{acc}=\text{\SI{5}{kV}, \SI{10}{kV}, \SI{15}{kV}, \SI{20}{kV}, and \SI{25}{kV}}}$.
The acquired spectra are shown in figure~\ref{fig:el_spectra}.
Only one pixel in the upper right of the detector was used in this particular measurement.
This pixel was chosen to avoid worsening the energy resolution due to light from the electron gun, see figure~\ref{fig:pm_fwhm_calib}.
At $V_\text{acc}=\SI{25}{kV}$, the main electron peak has an energy resolution of \SI{266}{eV} FWHM, which is close to the fundamental limit of \SI{240}{eV}~FWHM according to the Fano law in equation~\ref{eq:fano}.

\begin{figure}
    \centering
    \includegraphics{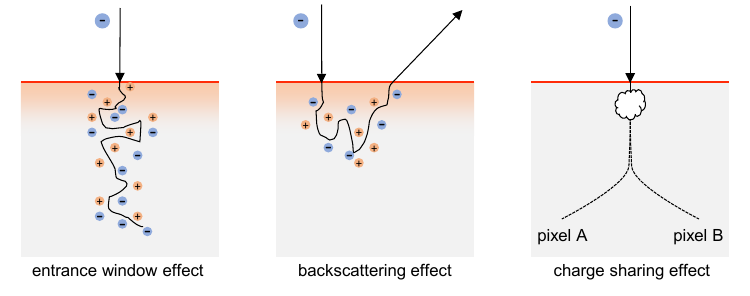}
    \caption{Illustration of three detector effects.
    The grey volume is the sensitive detector volume.
    At the entrance window effect, part of the charge cloud is lost due to the reduced charge collection efficiency close (a few tens of nanometers) to the entrance window (orange).
    The backscattering effect comes from electrons, which leave the detector due to scattering.
    Charge sharing, finally, happens at the border of two pixels, where the charge cloud can split into two.
    }
    \label{fig:electron_effects}
\end{figure}

Several physical effects characterize the observed spectral shape.
Three key effects, denoted as 1) entrance window effect, 2) backscattering effect, and 3) charge sharing effect, are depicted in figure~\ref{fig:electron_effects}.
The entrance window effect results in incomplete charge collection at the detector's entrance window, causing the main peak to be slightly asymmetric and shifted towards smaller energies. 
The backscattering effect leads to partial energy depositions, creating a broad low-energy shoulder extending from the energy threshold to the main peak.
Additionally, charge-sharing between neighboring pixels results in partial charge collection, further contributing to the low-energy shoulder.

Individual configurations of the measurement setup are necessary to characterize each effect in detail.
For example, on the one hand, a homogeneous illumination of the entrance window using the steering coils is advantageous to study the entrance window effect and charge sharing effect.
On the other hand, a high rate and small beam diameter are advantageous for studying the backscattering effect.
Detailed characterizations of the individual effects will be shown in other publications. Two publications are currently in preparation~\cite{spreng_24, siegmann_24}.

\section{Conclusion}

In the scope of this work, we have developed a novel electron gun for the characterization of silicon drift detectors~(SDD) with electrons.
The electron gun utilizes the thermionic emission of electrons from a heated tantalum filament.

We conducted several commissioning measurements to characterize the performance of the electron gun with a kinetic energy of up to \SI{25}{keV} and a rate of up to \SI{160E3}{cps}.
A steering coil system allows for the illumination of the area of one TRISTAN detector module ($40\times\SI{38}{mm^2}$).
Furthermore, a sufficient rate stability of \SI{5}{\percent} within 1.5 hours was observed.
We implemented an off-axis mounting and a light-shielding cup to prevent light of the glowing filament from hitting the detector.
This keeps the increase of the detector resolution due to light at an acceptable level. 

This work shows that the thermionic electron gun is well suited for characterizing the TRISTAN SDD detector with electrons.
Thus, the new electron gun is an essential tool for understanding electron-related detector effects, a key requirement for a sensitive search for keV-scale sterile neutrinos at the KATRIN experiment.

\section{Acknowledgements}
We acknowledge the support of the Deutsche Forschungsgemeinschaft DFG SFB-1258, the Ministry for Education and Research BMBF (05A23WO5), the Excellence Cluster ORIGINS, the Max Planck Research Group Program (MaxPlanck@TUM) and the Cusanuswerk. This project has received funding from the European Research Council (ERC) under the European Union Horizon 2020 research and innovation programme (grant agreement no. 852845).

\bibliographystyle{JHEP.bst}
\bibliography{references.bib}

\end{document}